%% file: lat04-abs252-jns.tex
\newif\ifpdf
\ifx\pdfoutput\undefined
        \pdffalse 
\else
        \pdfoutput=1
        \pdftrue
\fi

\documentclass[fleqn,twoside]{article}
\usepackage{espcrc2}

\ifpdf
	\usepackage[pdftex]{graphicx}
	\DeclareGraphicsExtensions{.jpg,.pdf,.mps,.png}
\else
	\usepackage[xdvi,dvips]{graphicx}
\fi

\input Macros.tex

\title{%
Leptonic decay constants
${f_{D_s}}$ and ${f_{D}}$ in three flavor
lattice QCD}

\author{%
J.N.\ Simone\address[FNAL]{%
Fermi National Accelerator Laboratory, P.O.\ Box 500, Batavia, IL 60510, USA}%
\thanks{talk presented by J.~Simone},
C.~Aubin\address[WashU]{%
Department of Physics, Washington University, St. Louis, MO 63130, USA},
C.~Bernard\addressmark[WashU],
C.~DeTar\address[Utah]{%
Physics Department, University of Utah, Salt Lake City, UT 84112, USA},
M.~di~Pierro\address[dePaul]{%
School of Computer Science, DePaul University, Chicago, IL 60604, USA},
A.X.~El-Khadra\address[UIUC]{%
Physics Department, University of Illinois, Urbana, IL 61801, USA},
Steven~Gottlieb\address[IU]{%
Department of Physics, Indiana University, Bloomington, IN 47405, USA},
E.B.~Gregory\address[Tucson]{%
Department of Physics, University of Arizona, Tucson, AZ 85721, USA},
U.M.~Heller\address[APS]{%
American Physical Society, One Research Road, Box 9000, Ridge, NY 11961, USA},
J.E.~Hetrick\address[UOP]{%
University of the Pacific, Stockton, CA 95211, USA},
A.S.~Kronfeld\addressmark[FNAL],
P.B.~Mackenzie\addressmark[FNAL],
D.P.~Menscher\addressmark[UIUC],
M.~Nobes\address[SFU]{%
Physics Department, Simon Fraser University, Burnaby, BC, Canada},
M.~Okamoto\addressmark[FNAL],
M.B.~Oktay\addressmark[UIUC],
J.~Osborn\addressmark[Utah],
R.~Sugar\address[UCSB]{%
Department of Physics, University of California, Santa Barbara, CA, 93106, USA},
D.~Toussaint\addressmark[Tucson],
and H.D.~Trottier\addressmark[SFU]
}

\newcommand{\FIGextrapFullQCDline}{%
\begin{figure}[htb]
\ifpdf
   \includegraphics[clip=true,width=1.0\columnwidth]{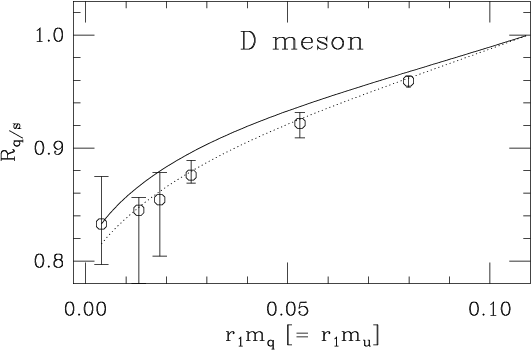}
\else
   \includegraphics[clip=true,width=1.0\columnwidth]{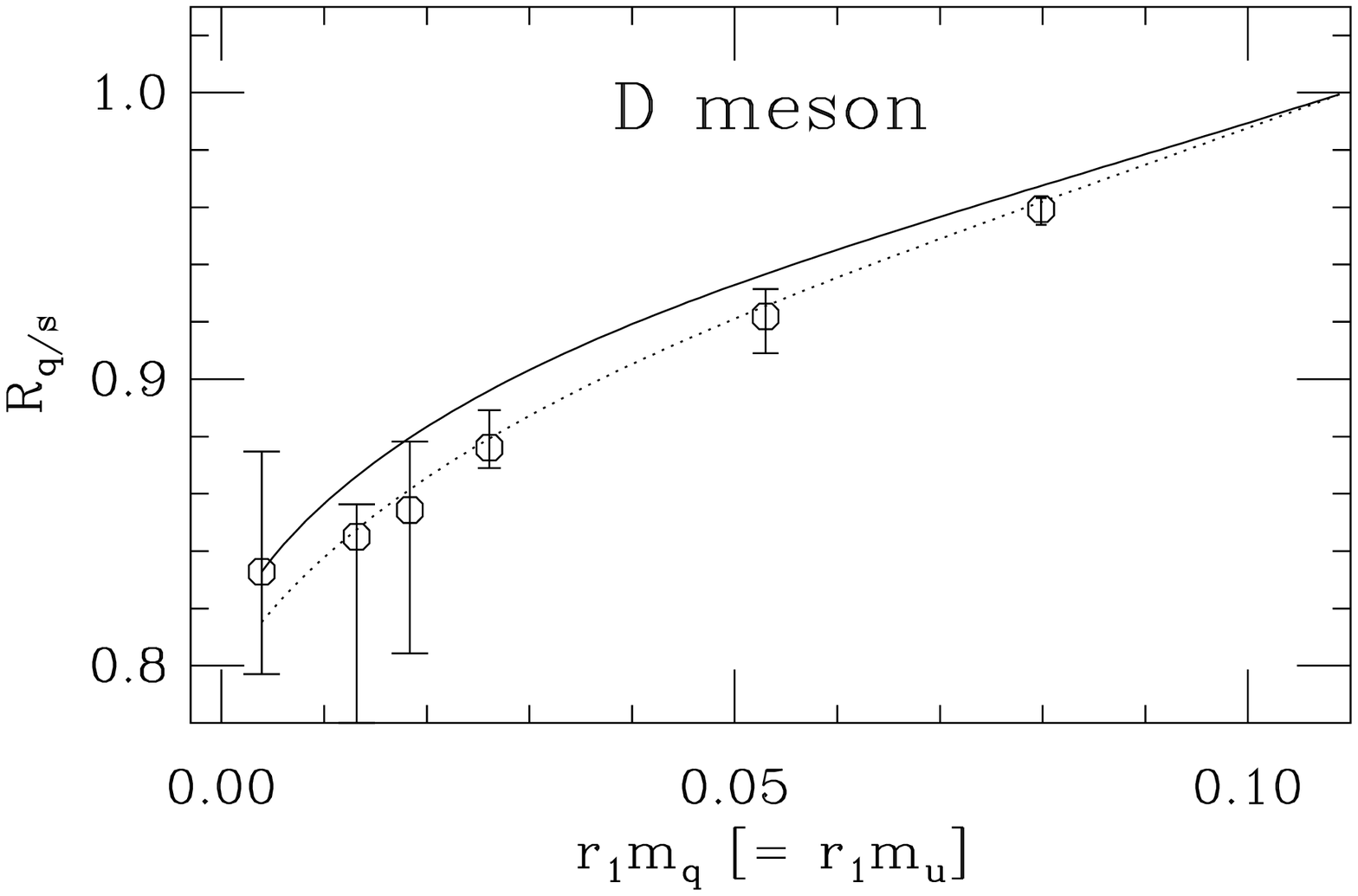}
\fi
\mbox{\relax}\vspace{-4.0ex}
\caption{The chiral extrapolation of the ratio $\fSqrtM{D}/\fSqrtM{D_s}$.
The  ratio is  determined by the  solid curve  where staggering
effects   have  been   removed.  The   dashed  curve   includes  these
discretization  effects.  }
\label{figure:extrapFullQCDline}
\end{figure}}

\newcommand{\FIGextrapDs}{%
\begin{figure}[htb]
\ifpdf
   \includegraphics[clip=true,width=1.0\columnwidth]{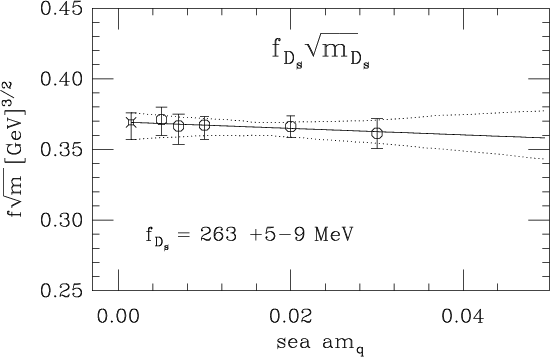}
\else
   \includegraphics[clip=true,width=1.0\columnwidth]{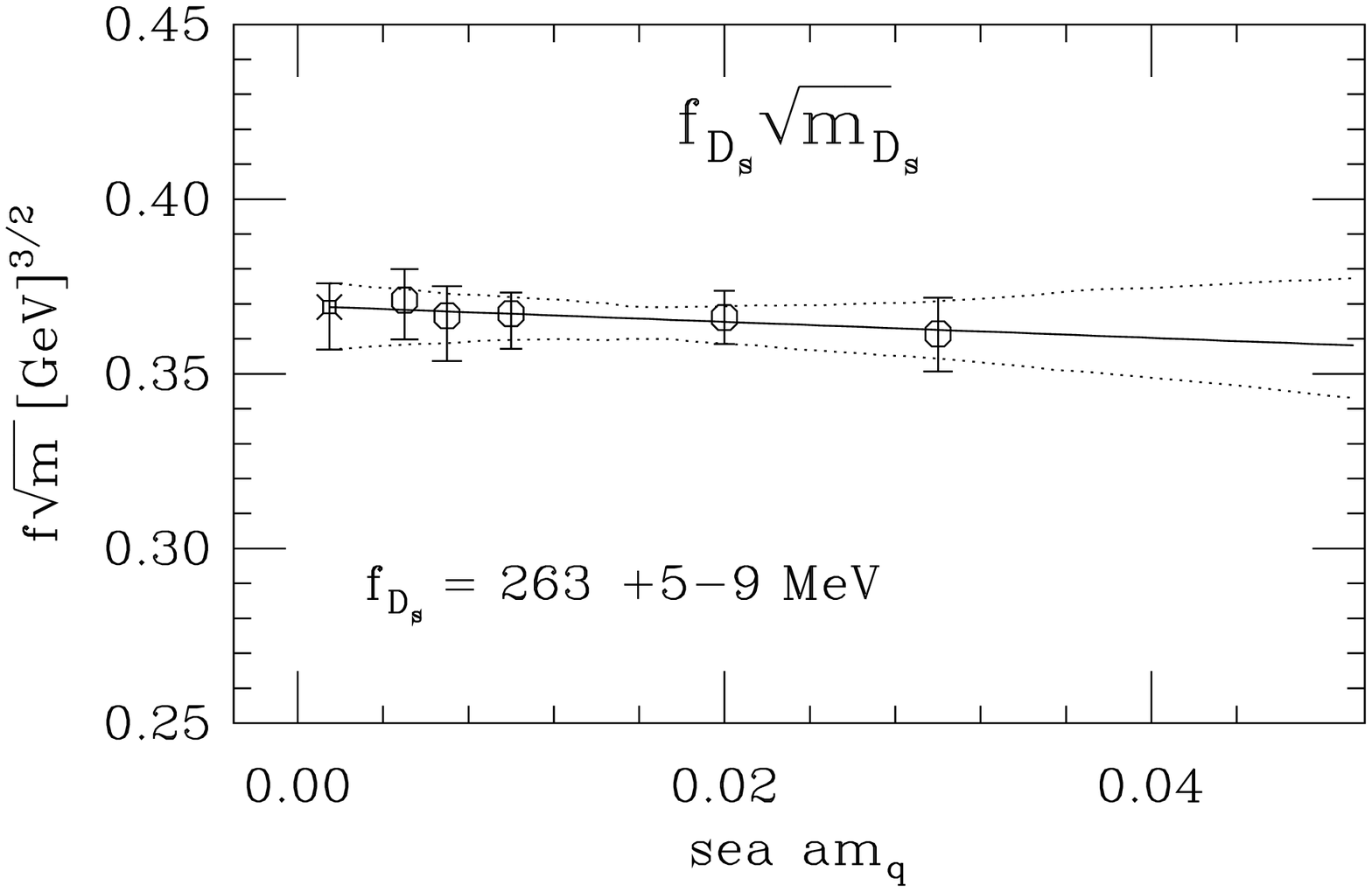}
\fi
\mbox{\relax}\vspace{-4.0ex}
\caption{Extrapolation in the light sea quark mass for $\fSqrtM{D_s}$.
The curves show a linear
fit (soild) and the 68\% confidence level statistical error bounds (dotted).
}
\label{figure:extrapDs}
\end{figure}}

\newcommand{\errorBudget}{%
\begin{table}[htb]
\caption{Error budget as percentage of each quantity.}
\label{table:errorBudget}
\begin{center}
\begin{tabular}{lccc}
source               &${R_{d/s}}$  &${\fSqrtM{D_s}}$  &${\fSqrtM{D}}$ \\ \hline
{stat.+extrap.}
                     &4.7                 &3.3                      &6.2  \\
{HQ matching}
                    &$<$1                 &7                        &7   \\
{LQ discret.}
                    &4                    &4                        &4    \\
{${m_{c}}$ det.}
                    &$<$1                 &4                        &4    \\
{val. ${m_s}$, ${m_d}$}
                    &2                    &1                        &2.2  \\
{${a}$ \& sea quark}
                    &$<$1                 &2                        &2 
\end{tabular}
\end{center}
\end{table}}

\begin{document}

\begin{abstract}
We determine the leptonic decay constants ${\f{D_s}}$ and ${\f{D}}$ in
three  flavor  unquenched   lattice  QCD.   We  use  $O(a^2)$-improved
staggered  light  quarks  and  $O(a)$-improved  charm  quarks  in  the
Fermilab heavy  quark formalism.  Our preliminary  results, based upon
an   analysis  at   a   single  lattice   spacing,   are  $\f{D_s}   =
263\Aerr{5}{9}\pm24\;\MeV$   and  $\f{D}=225\Aerr{11}{13}\pm21\;\MeV$.
In each case, the first reported error is statistical while the second
is the combined systematic uncertainty.
\vspace{0pc}
\end{abstract}

\maketitle

\section{INTRODUCTION}

The  leptonic decay constants  $\f{B_s}$ and  $\f{B}$ are  critical in
testing  the flavor sector  of the  Standard Model.   Reliable lattice
calculations are of fundamental  importance since the determination of
these decay constants remains beyond the reach of experiment.

Precise experimental  determinations of the  decay constants $\f{D_s}$
and  $\f{D}$ and  the semileptonic  decays $\DtoPi$  and  $\DtoK$ will
result from the high-statistics charm  program of CLEO-c.  Comparing
these experimental results  with lattice results will serve  both as a
critical  check  of  lattice methods  for  charm  and  as a  means  of
assessing  the  reliability of  lattice  calculations  for the  bottom
quark.

This  work calculates  the  leptonic decay  constants ${\f{D_s}}$  and
${\f{D}}$   using  $O(a^2)$-improved   staggered   light  quarks   and
$O(a)$-improved  charm quarks  in the  Fermilab heavy  quark formalism
\cite{El-Khadra:1996mp}.  It was done  with three flavors of light sea
quarks.   The staggered  fermion  action for  the  light quarks  makes
possible calculations  at lighter  quarks masses than  previously used
\cite{Bernard:2004kz},    allowing   a   better    controlled   chiral
extrapolation.  We  use the  results of staggered,  partially quenched
chiral  perturbation  theory  (S$\chi$PT)  in  performing  the  chiral
extrapolations.

\section{\boldmath${\fSqrtM{D}/\fSqrtM{D_s}}$ RATIO}

We  use the  MILC collaboration  Asqtad gauge  ensembles  with lattice
spacing $a  \approx1/8 $  fm \cite{Aubin:2004wf}.  Each  ensemble has
one sea quark  flavor of mass, $m_s$, approximating  the strange quark
and two flavors, degenerate in mass, $m_u$.  The five ensembles we use
have light flavors in the range  $0.1 m_s \leq m_u \leq 0.6 m_s$.  For
each ensemble, decay constants were computed at twelve logarithmically
spaced values of  the valence quark mass in the  range $0.1 m_s\le m_q
\le m_s$.  In all, decay  constants were computed for  sixty partially
quenched $(m_q,m_u)$ combinations.

The chiral expansion for the decay constants is known at leading order
in both  the heavy quark  expansion and staggered  chiral perturbation
theory \cite{Aubin:2004xd}. We take the parameterization
\begin{eqnarray}
\label{eqn:chiralExtrapolationParam}  \nonumber
R_{q/s}&=&1 + a_0 ( \Delta f_q - \Delta f_s ) +
(m_q - m_s)\times
\\
&&( a_1 +
a_2 \tilde{m} + a_3 m_q + a_4 m_s+\ldots)
\end{eqnarray}
for the ratio $R_{q/s}=\fSqrtM{D_q}/\fSqrtM{D_s}$ in our chiral
extrapolation.  For staggered quarks, the chiral log terms $\Delta
f_x$ contain discretization effects from taste violations in the
pseudoscalar masses as well as explicit taste violation terms.  We
include quark mass terms up to $O(m^2)$ with the constraint
$a_4=a_3-a_1^2$. The sum of sea quark masses is $\tilde{m}\equiv
2m_u+m_s$.

The  parameters  $a_j$  in Eq.~(\ref{eqn:chiralExtrapolationParam})  are
determined in a fully  covariant $\chi^2$ minimization using all sixty
partially  quenched decay constant  results.  Pseudoscalar  masses and
coefficients of the  explicit taste violation terms were  fixed to the
values   determined    in   an   analysis   of    the   light   mesons
\cite{Aubin:2004fs}.  Bayesian priors were input for each $a_j$.  With
$\xi=1/(4\pi f)^2$, the prior for $a_0$ is $-0.5\xi(1+3g^2)(1\pm0.30)$
with $g^2\approx0.35$.  Other priors are  $0\pm1$ in units of $2\xi s$
where  $s$ is  the slope  relating the  quark mass  and the  pion mass
squared.

\FIGextrapFullQCDline

The fit,  shown in Fig.~\ref{figure:extrapFullQCDline},  has $\chi^2/{
\rm  dof}=0.2$.  The  extrapolation according  to S$\chi$PT  is shown
along the ``full  QCD'' direction where the valence  quark mass equals
the  sea  quark  mass.   The  parameters  in  this  extrapolation  are
determined  using  all  sixty  partially-quenched points.   We  obtain
$R_{d/s} =  0.833\Aerr{0.042}{0.036}$ in the chiral  limit.  The error
is  the  combined  uncertainty  from statistics  and  parameter  prior
estimates  for the  extrapolation function.   Systematic  effects from
matching the lattice theory to QCD, the lattice spacing and the tuning
of   charm  quark  mass   mostly  cancel   in  the   ratio.   Residual
discretization  effects   from  light  quarks,  not   removed  by  the
extrapolation procedure,  are estimated to be $4\%$.   Our estimate is
based  on   the  size  of   known  taste-breaking  effects   shown  in
Fig.~\ref{figure:extrapFullQCDline}  and is similar  to taste-breaking
effects found in $f_K$ and $f_\pi$ \cite{Aubin:2004fs}.  The ratio was
determined using the nominal strange  quark mass rather than the tuned
value which leads to an uncertainty of $2\%$.

\section{\boldmath ${\fSqrtM{D_s}}$ DETERMINATION}

\FIGextrapDs

We obtain  $\fSqrtM{D_s}$ by first interpolating to  the tuned valence
$m_s$  value  obtained  from  the  light  mesons  on  the  same  gauge
ensembles.     Then,   a   mild    sea   quark    extrapolation   (see
Fig.~\ref{figure:extrapDs})   is  needed   to  obtain   $\f{Ds}$.   We
extrapolate linearly to the tuned $\hat{m}=(m_u+m_d)/2$ value obtained
from  the light  mesons \cite{Aubin:2004fs}.  The  combined statistical
and extrapolation  error is $3.3\%$.  The uncertainty  from the tuning
of $m_s$ and  $\hat{m}$ is $1\%$.  The tuning of  the charm mass leads
to  a   $4\%$  error.   The  lattice  spacing   uncertainty  is  $2\%$
\cite{Davies:2003ik}.   The dominant  systematic uncertainty,  7\%, is
from the mismatch between the  lattice theory and QCD, as discussed in
Ref.~\cite{Aubin:2004ej}.  Our final  results will include an improved
estimate  of this  uncertainty  incorporating results  from finer  and
coarser lattice spacings, which are now in progress.

\section{RESULTS}

Statistical   and   systematic   uncertainties   are   summarized   in
Table~\ref{table:errorBudget}.  Our estimates  of heavy quark matching
effects and  light quark discretization  effects are based  on results
from a single lattice spacing.  We will refine our error estimates and
update  our  results  once  decay constants  from  additional  lattice
spacings  are  known. The  heavy  quark  matching  uncertainty can  be
reduced by including the higher order matchings for the action and the
currents
\cite{Nobes:2003nc,Oktay:2003gk}.

\errorBudget

Combining  in   quadrature  the  systematic   uncertainties shown in
Table~\ref{table:errorBudget}, we find our preliminary results:

\begin{minipage}{\textwidth}
\begin{eqnarray*}
\frac{\fSqrtM{D_s}}{\fSqrtM{D}} &= &1.20\pm0.06\pm0.06\;\Comma \\
f_{D_s} &= &263\Aerr{5}{9}\pm24 \quad\textrm{MeV}\;\Comma\\
f_{D}   &= &225\Aerr{11}{13}\pm21 \quad\textrm{MeV}
\;\FullStop
\end{eqnarray*}
\end{minipage}

\section*{ACKNOWLEDGMENTS}

We acknowledge  support through  the DOE and  NSF high  energy physics
programs.   We thank  the DOE  SciDAC program  and  Fermilab Computing
Division for  support. Fermilab  is operated by  Universities Research
Association Inc.  under contract with the United  States Department of
Energy.

\end{document}

%% file: Macros.tex
\newcommand{\simgt}%
{\mathrel{\lower0.5ex\hbox{\ensuremath{\buildrel>\over\sim}}}}
\newcommand{\simlt}%
{\mathrel{\lower0.5ex\hbox{\ensuremath{\buildrel<\over\sim}}}}
\newcommand{\BarLongRightArrow}%
{\ensuremath{\mathrel\vert\joinrel\Longrightarrow}}

\newcommand{\FullStop}{.}
\newcommand{\Comma}{,}

\newcommand{\MeV}{\textrm{MeV}}

\newcommand{\DtoPi}{\ensuremath{D \to \pi\ell\nu}}
\newcommand{\DtoK}{\ensuremath{D \to K\ell\nu}}

\newcommand{\f}[1]{\ensuremath{f_{#1}}}
\newcommand{\fSqrtM}[1]{\ensuremath{\f{#1}\sqrt{m_{#1}}}}

\newcommand{\Aerr}[2]{\ensuremath{\vphantom{0}^{+{#1}}_{-{#2}}}}

